# Investigation of in-plane anisotropy of $c$-axis magnetoresistance for BiCh$_2$-based layered superconductor NdO$_{0.7}$F$_{0.3}$BiS$_2$

Kazuhisa Hoshi[1], Kenta Sudo[2], Yosuke Goto[1], Motoi Kimata[2], Yoshikazu Mizuguchi[1]

1 *Department of Physics, Tokyo Metropolitan University, 1-1, Minami-osawa, Hachioji 192-0397, Japan*
2 *Institute for Materials Research, Tohoku University, 2-1-1, Katahira, Sendai 980-8577, Japan*

Abstract

We have investigated the in-plane anisotropy of the $c$-axis magnetoresistance (MR) in both superconducting and normal states of single crystals of NdO$_{0.7}$F$_{0.3}$BiS$_2$ under in-plane magnetic fields. In the superconducting states of NdO$_{0.7}$F$_{0.3}$BiS$_2$, four-fold-symmetric in-plane anisotropy of the $c$-axis MR was observed below the superconducting transition temperature. Since the crystal structure of NdO$_{0.7}$F$_{0.3}$BiS$_2$ is tetragonal, the rotational symmetry in the superconducting state is preserved in the present compound. This result is clearly different from the previous report observed in LaO$_{0.5}$F$_{0.5}$BiSSe single crystals, where the in-plane MR in the superconducting state shows two-fold symmetry. On the other hand, in the normal states of NdO$_{0.7}$F$_{0.3}$BiS$_2$, two-fold symmetric MR with a small amplitude was observed. The possible origin of the two-fold-symmetric behavior was discussed with the presence of local structural disorder in the conducting plane of BiCh$_2$-based compounds.

# 1. Introduction

The BiCh$_2$-based (Ch: S, Se) superconductor family was discovered in 2012 [1–3] and has been extensively studied because the layered crystal structure resemble those of the cuprate and Fe-based high-transition-temperature (high-$T_c$) superconductors [4,5]. The typical REOBiCh$_2$ (RE: Rare earth) system has a layered structure composed of the alternate stacks of REO insulating layers and BiCh$_2$ conducting layers [2,3]. Fluorine substitutions for the oxygen site lead to electron carriers in the BiCh$_2$ layers, which results in the emergence of superconductivity at low temperatures. The superconducting mechanisms in BiCh$_2$-based superconductors have been extensively studied from both theoretical and experimental aspects. From theoretical studies, the possibilities of conventional *s*-wave, extended *s*-wave, *d*-wave, and *g*-wave states including weak topological superconductivity states have been investigated for BiCh$_2$-based superconductors [6-12]. Many experimental studies, such as thermal conductivity measurement, magnetic penetration depth, specific heat, and muon spin resonance ($\mu$SR) propose conventional *s*-wave superconducting mechanism in BiCh$_2$-based superconductor [13-15]. However, angle resolved photoemission spectroscopy (ARPES) measurement indicated the presence of anisotropic superconducting gap in NdO$_{0.71}$F$_{0.29}$BiS$_2$ [17]. Also, the anomalous superconducting gap states have been reported by scanning tunneling spectroscopy (STS) for Nd(O,F)BiS$_2$ [18]. In addition, the absence of isotope effect on $T_c$ was confirmed in LaO$_{0.6}$F$_{0.4}$Bi(S, Se)$_2$ by using $^{76}$Se and $^{80}$Se isotopes [19]. These results suggest possibility of unconventional superconductivity in BiCh$_2$-based superconductors. Thus, the discussion about the mechanisms of superconductivity in BiCh$_2$-based superconductors is still controversial.

Recently, electronic nematicity has been a hot topic in the field of superconductivity. Electronic nematicity results in electronic states which break rotational symmetry of the crystal lattice. This electronic nematicity has been particularly studied in Fe-based superconductors [20, 21]. Moreover, the nematicity in superconducting state (called nematic superconductivity) has been reported in the candidates of topological superconductor; typical examples are doped Bi$_2$Se$_3$ [22-25]. In the nematic superconductivity, the rotational symmetry of superconducting gap amplitude breaks underlying symmetry of the crystal lattice. More recently, we have observed two-fold symmetric in-plane anisotropy of the *c*-axis magnetoresistance (MR) in the superconducting states of LaO$_{1-x}$F$_x$BiSSe single crystals ($x$ = 0.1 and 0.5) [26,27]. Since the crystal structure is tetragonal (P4/*nmm*) with four-fold symmetry in the *ab*-plane, the observed two-fold-symmetry in the superconducting states is resulting after breaking rotational symmetry of the crystal lattice. In addition, from a structural phase diagram of LaO$_{1-x}$F$_x$BiSSe, we confirmed that a structural transition from tetragonal to monoclinic at low temperatures is completely suppressed by 4%-F substitution for the O site, i.e., $x > 0.04$ [28]. However, the universality of symmetry-breaking MR in various type of BiCh$_2$ superconducting family is still an open question, and thus further investigation including different compositions would be helpful to clarify the overall features of MR symmetry in the superconducting states. In this study, we show that in-plane anisotropy of the *c*-axis MR for BiCh$_2$-based superconductor NdO$_{0.7}$F$_{0.3}$BiS$_2$ single crystals in both superconducting and normal states. We found that four-fold symmetry of the *c*-axis MR in the superconducting states. This result is clearly different from those observed for LaO$_{1-x}$F$_x$BiSSe in which two-fold symmetry was observed in superconducting states [26,27].

## 2. Experimental details

NdO$_{0.7}$F$_{0.3}$BiS$_2$ single crystals were grown by a high-temperature flux method in an evacuated quartz tube [29]. Polycrystalline samples of NdO$_{0.7}$F$_{0.3}$BiS$_2$ were prepared by the solid-state-reaction method using powders of Nd$_2$S$_3$ (99.9%), Bi$_2$O$_3$ (99.999%), and BiF$_3$ (99.9%) and grains of Bi (99.999%) and S (99.999%). A mixture of the starting materials with a nominal ratio of NdO$_{0.7}$F$_{0.3}$BiS$_2$ was mixed, pressed into a pellet, and annealed at 700 ºC for 15 h in an evacuated quartz tube. The obtained sample was grounded, mixed, pelletized, and heated at same temperature. The polycrystalline powder of NdO$_{0.7}$F$_{0.3}$BiS$_2$ (0.5 g) was mixed with CsCl/KCl flux (3.0 g), and the mixture was sealed into an evacuated quartz tube. The tube was heated at 800 ºC for 15 h and slowly cooled to 600 ºC with a rate of -1.0 ºC /h, followed by furnace cooling to room temperature. At room temperature, the quartz tube was opened under air atmosphere, and the product was filtered and washed with pure water. Single crystals were analyzed by scanning electron microscopy (SEM) TM3030 (Hitachi high-tech). As shown in Fig. 1(a), plate-like crystals were obtained. The plate surface with a square shape is corresponding to the *ab*-plane of tetragonal NdO$_{0.7}$F$_{0.3}$BiS$_2$ [3,29]. The chemical composition of the obtained crystals was investigated by energy-dispersive X-ray spectroscopy (EDX) on the SEM. The average ratio of the constituent elements (except for O and F) was Nd : Bi : S = 1 : 1.0 : 2.1, which was normalized by the Nd value. The analyzed atomic ratio is almost consistent with the nominal composition.

The MR measurement was performed using a superconducting magnet at the high field laboratory of Institute for Materials Research (IMR), Tohoku University. To precisely control the magnetic field direction, a two-axes rotational probe was used. Figure 1(b) shows a schematic image of the terminal configuration, which was made using Au wires and Ag pastes, for the *c*-axis resistivity measurements.

## 3. Results and discussion

Figure 1(c) shows the temperature (*T*) dependence of the *c*-axis resistivity ($\rho_c$) of NdO$_{0.7}$F$_{0.3}$BiS$_2$. The inset of Fig.1 (c) shows enlarged temperature dependence of the $\rho_c$ ($\rho_c$-*T*) around superconducting transition. The $T_c^{onset}$ is 5.8 K, and $T_c^{zero}$ is 5.2 K. A small upturn behavior on the $\rho_c$-$T$ is observed at low temperature. Similar upturn behavior is observed for some layered superconductors having two-dimensional transport characteristics, such as cuprate and Fe-based superconductors [30,31].

As shown in Fig. 2(a), $\theta$ and $\phi$ are defined. $\theta$ is measured from the $c$-axis, and $\phi$ is azimuth angle measured from the $a$-axis to the $b$-axis, while the $a$-axis and the $b$-axis are equivalent in a tetragonal crystal. Note that the charge current is always perpendicular to the magnetic field when the magnetic field is in the conducting plane since the $c$-axis resistivity was measured. Figure 2(b) shows the $\theta$ angle dependence of the $c$-axis electrical resistivity at $\phi = 180°$, $B = 14$ T, and $T = 2.7$ K. As shown in Fig. 2(b), $\rho_{B \parallel ab}$, which was defined as the MR where the magnetic field is exactly parallel to the $ab$-plane, was estimated and used to investigate the in-plane anisotropy.

Figure 2(c) shows the $\phi$ angle dependences of the $\rho_{B \parallel ab}$ at $B = 14$ T measured at temperatures from 10.0 to 2.5 K. Four-fold symmetry of the $\rho_{B \parallel ab}$ was observed in the superconducting states, which is trivial behavior because $ab$-plane of $NdO_{0.7}F_{0.3}BiS_2$ single crystal has a tetragonal square Bi-S plane with four-fold symmetry. The result suggests that rotational symmetry in the crystal lattice is preserved when we see the in-plane anisotropy of the superconducting properties in $NdO_{0.7}F_{0.3}BiS_2$. Similar four-fold-symmetric in-plane MR in the superconducting states was observed for a crystal taken from a different batch (See supplementary information). This fact is clearly different from the two-fold symmetric MR observed in the superconducting states of $LaO_{1-x}F_xBiSSe$ single crystals [26,27].

Although we cannot conclude the origins of the switching between the two-fold-symmetric and four-fold-symmetric superconducting properties in $BiCh_2$-based superconductors so far, here we briefly discuss the possible origins of the different behaviors of the in-plane anisotropy for $NdO_{0.7}F_{0.3}BiS_2$ and $LaO_{1-x}F_xBiSSe$ to encourage further theoretical and experimental studies on those systems. First, in $LaO_{1-x}F_xBiSSe$, two-fold-symmetric behavior was observed for $x = 0.1$ and 0.5. Since the carrier concentration and the Fermi surface topology could be different each other for those F concentrations, one can assume that the emergence of two-fold-symmetric behavior is not affected by the carrier concentration. Then, we compare the constituent elements in the superconducting layer in $NdO_{0.7}F_{0.3}BiS_2$ and $LaO_{1-x}F_xBiSSe$. $NdO_{0.7}F_{0.3}BiS_2$ contains Bi-S superconducting planes, while $LaO_{1-x}F_xBiSSe$ contains nearly Bi-Se superconducting planes, according to the structural analysis of $LaO_{1-x}F_xBiSSe$ [31]. The Se substitution for S at the superconducting plane should enhance spin-orbit interaction, and the spin-orbit interaction is known as an important parameter to understand the electronic states and superconducting characteristics in the $BiCh_2$-based systems [33-36]. The importance of spin-orbit interaction is also pointed out in the doped $Bi_2Se_3$ systems, which show nematic superconducting behavior. Thus, the stronger spin-orbit interaction in the conducting plane of LaO$_{1-x}$F$_x$BiSSe might be crucial for the symmetry breaking MR in this compound. To conclude the origins for four-fold and two-fold-symmetric anisotropies of the MR in the superconducting states in NdO$_{0.7}$F$_{0.3}$BiS$_2$ and LaO$_{1-x}$F$_x$BiSSe, further studies are desired.

On the normal state properties, we observed two-fold-symmetric in-plane anisotropy of MR, as shown in Fig. 3(a). For comparison, the normal-state MR anisotropy for LaO$_{0.5}$F$_{0.5}$BiSSe is shown in Fig. 3(b), in which four-fold symmetry is dominant. Note that the amplitude of those signals is very small as shown in Fig. 2(c). We consider that those anisotropy observed in the normal states is not directly linked to those in the superconducting states for both systems. Basically, the BiCh$_2$-based compounds possess in-plane structural disorder due to the presence of Bi lone-pair electrons [37-40]. From extended X-ray absorption fine structure studies, in-plane local structures were explained using two different Bi-Ch bonds. However, the local disorder can be suppressed by chemical pressure effects [37,38], which is higher in a Bi-Se plane than in a Bi-S plane [40]. Therefore, we expect disordered local structure in NdO$_{0.7}$F$_{0.3}$BiS$_2$ than LaO$_{1-x}$F$_x$BiSSe. Although the samples used in the experiments are single crystals, such intrinsic local disorder (formation of local domains) would exist in both systems, and that should be more prominent in the transport properties of NdO$_{0.7}$F$_{0.3}$BiS$_2$ than LaO$_{1-x}$F$_x$BiSSe.

## 4. Summary

We have investigated the in-plane anisotropy of the *c*-axis MR in both superconducting and normal states of single crystals of NdO$_{0.7}$F$_{0.3}$BiS$_2$ under in-plane magnetic fields. In the superconducting states, four-fold-symmetric in-plane anisotropy of MR was observed below the $T_c$. Since the crystal structure of NdO$_{0.7}$F$_{0.3}$BiS$_2$ is tetragonal, the results suggest no symmetry breaking in the in-plane anisotropy of MR in the superconducting states. This is clearly different from those observed for LaO$_{0.5}$F$_{0.5}$BiSSe single crystals, in which two-fold-symmetric anisotropy of MR was observed in its superconducting states. We consider that the switching between two-fold-symmetric and four-fold-symmetric anisotropy of MR in the superconducting properties of those systems is related to the constituent elements at the superconducting planes, which is expected to modify the spin-orbit interactions. In the normal states of NdO$_{0.7}$F$_{0.3}$BiS$_2$, two-fold symmetric MR with a small amplitude was observed, while four-fold-symmetric anisotropy was observed in the normal states of LaO$_{0.5}$F$_{0.5}$BiSSe. We consider that the normal-state in-plane anisotropy of MR in NdO$_{0.7}$F$_{0.3}$BiS$_2$ is caused by local structural instability that is commonly observed in BiCh$_2$-based compounds and tunable by chemical pressures.

**Acknowledgement**

The authors thank O. Miura for experimental support. This work was partly supported by Collaborative Research with IMR, Tohoku Univ. (proposal number: 18H0006, 19H0023), Grants-in-Aid for Scientific Research (Nos. 15H05886, 16H04493, 18KK0076, and 19K15291) and the Advanced Research Program under the Human Resources Funds of Tokyo (Grant Number: H31-1).

**Figures**

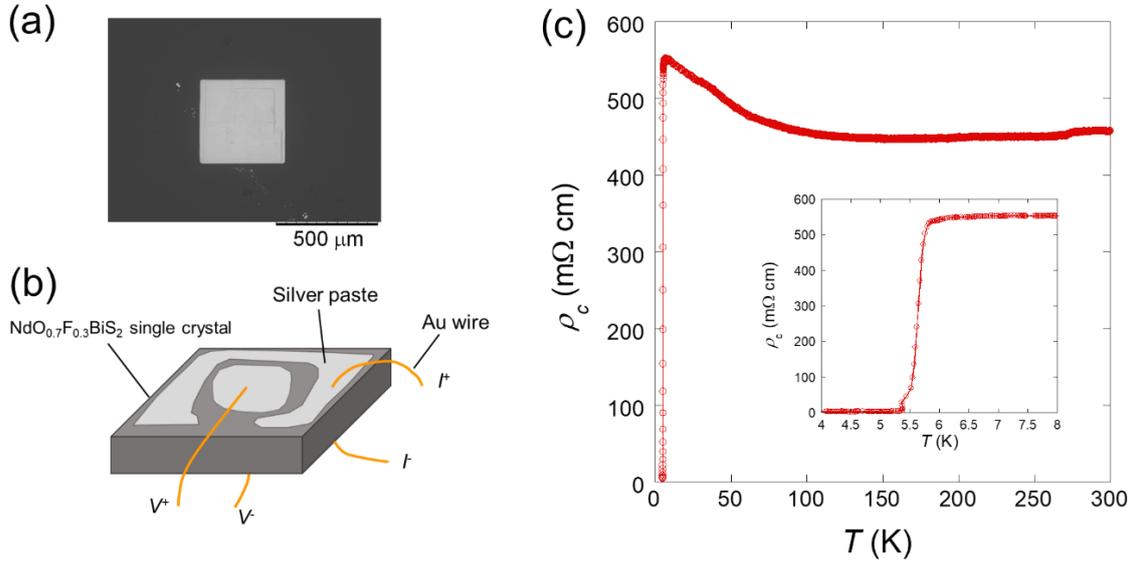

Fig. 1. (a) SEM image of a NdO$_{0.7}$F$_{0.3}$BiS$_2$ single crystal. (b) Schematic image of the terminal configuration for the *c*-axis resistivity measurement performed on NdO$_{0.7}$F$_{0.3}$BiS$_2$ single crystals. (c) Temperature dependence of the *c*-axis resistivity for NdO$_{0.7}$F$_{0.3}$BiS$_2$. The inset shows the enlarged resistivity data near a superconducting transition.

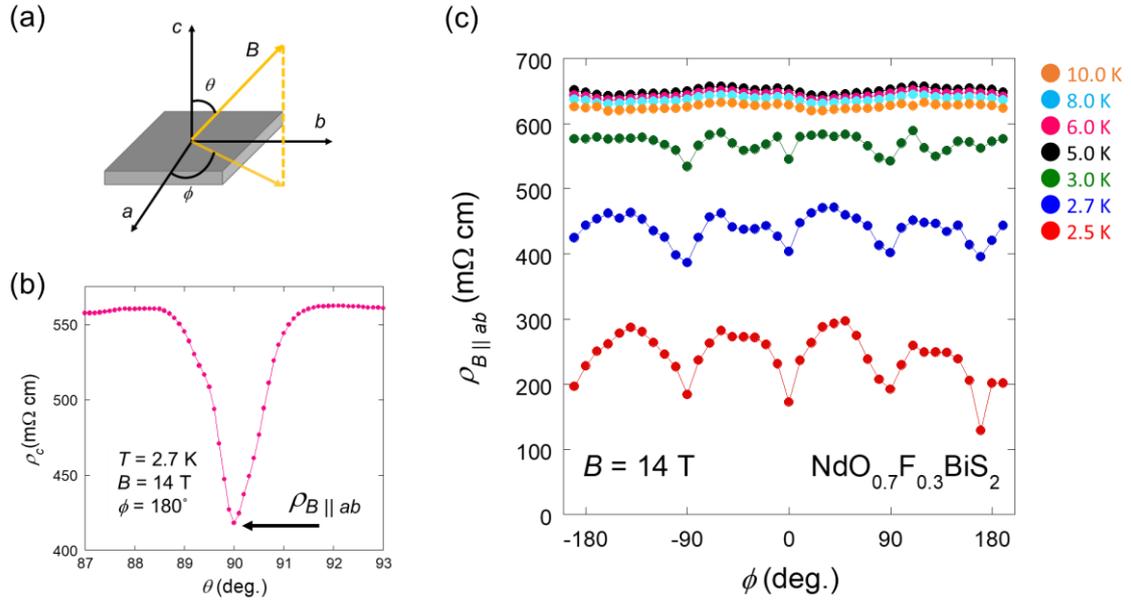

Fig. 2. (a) Schematic image of the rotation angles for the in-plane anisotropy measurement. (b) $\theta$ angle dependence of the $c$-axis resistivity for NdO$_{0.7}$F$_{0.3}$BiS$_2$ measured at the condition of $\phi=180°$, $B = 14$ T, and $T = 2.7$ K. (c) $\phi$ angle dependences of the $\rho_{B \parallel ab}$ for NdO$_{0.7}$F$_{0.3}$BiS$_2$ measured at $B = 14$ T and $T = 2.5$–10.0 K.

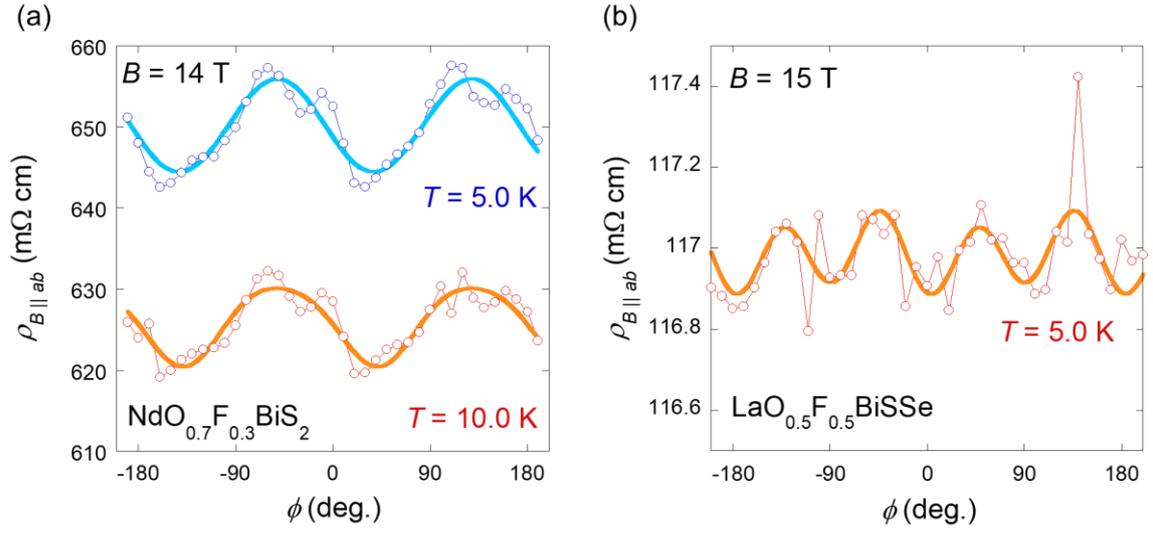

Fig. 3. (a) $\phi$ angle dependences of the normal-state MR for NdO$_{0.7}$F$_{0.3}$BiS$_2$ measured at $B$ =14 T in normal states at $T$ = 5.0–10.0 K. (b) $\phi$ angle dependences of the normal-state MR for LaO$_{0.5}$F$_{0.5}$BiSSe measured at $B$ =15 T and at $T$ = 5.0 K. The plotted data was fitted by the function of $A\sin(2\phi + \alpha) + B\sin(4\phi + \beta) + C$.

# Supplementary information

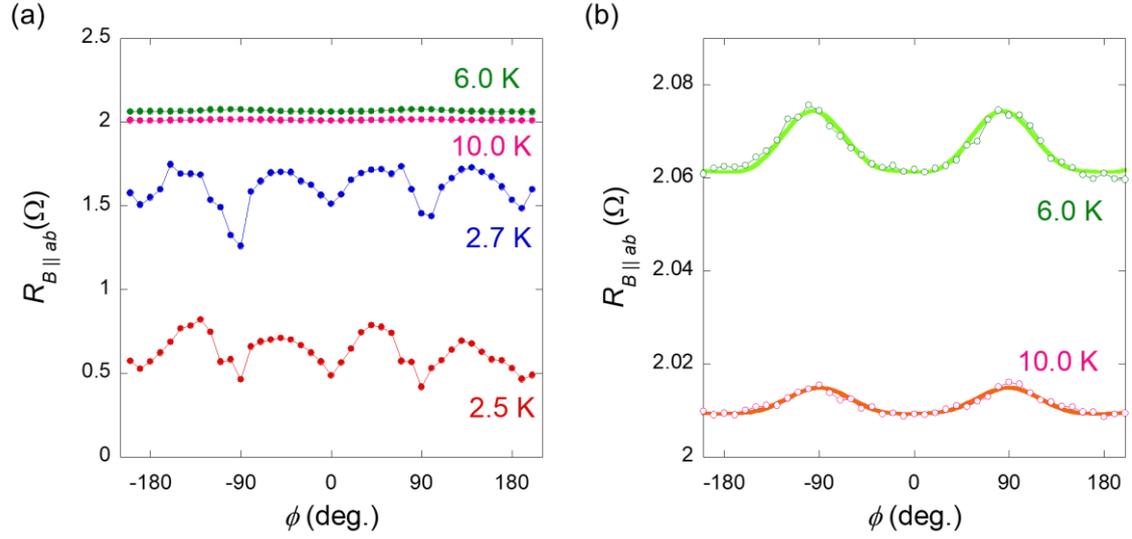

Fig. S1. (a) $\phi$ angle dependences of the superconducting-state MR for NdO$_{0.7}$F$_{0.3}$BiS$_2$ obtained from a different batch measured at $B =14$ T and $T = 2.5$–$10.0$ K. (b) $\phi$ angle dependences of the MR for NdO$_{0.7}$F$_{0.3}$BiS$_2$ measured at $B =14$ T and $T = 6.0$ and $10.0$ K. The plotted data was fitted by the function of $A\sin(2\phi + \alpha) + B\sin(4\phi + \beta) + C$.